\documentclass[11pt]{article}
\usepackage{a4wide}
\usepackage{amsmath, amssymb, amscd, amsthm, amsfonts}
\usepackage[utf8]{inputenc}
\usepackage{graphicx}
\usepackage{hyperref}
\usepackage{cite}
\usepackage[dvipsnames]{xcolor}

\newcommand{\Tr}{{\rm Tr}}

\newcommand{\be}{\begin{equation}}
\newcommand{\ee}{\end{equation}}
\newcommand{\ba}{\begin{align}}
\newcommand{\ea}{\end{align}}
\newcommand{\ra}{\rangle}
\newcommand{\la}{\langle}

\begin{document}
{\renewcommand{\thefootnote}{\fnsymbol{footnote}}
		
\begin{center}
{\LARGE IKKT thermodynamics and early universe cosmology} 
\vspace{1.5em}

Samuel Laliberte$^{1}$\footnote{e-mail address: {\tt samuel.laliberte@mail.mcgill.ca}} and
Suddhasattwa Brahma$^{2}$\footnote{e-mail address: {\tt suddhasattwa.brahma@gmail.com}}
\\
\vspace{1.5em}
$^1$Department of Physics, McGill University, Montr\'{e}al, QC, H3A 2T8, Canada\\
$^2$Higgs Centre for Theoretical Physics, School of Physics and Astronomy,\\ University of Edinburgh, Edinburgh EH9 3FD, UK\\

\vspace{1.5em}
\end{center}
}
	
\setcounter{footnote}{0}

\begin{abstract}
	\noindent Matrix theory is a proposed non-perturbative definition of superstring theory in which space is emergent.  Recently, it was shown that space-time can emerge with a scale-invariant spectrum of cosmological perturbations which is sourced by thermal fluctuations of the BFSS model at finite temperature.  Inspired by these results, we begin a study of the IKKT model at finite temperature.  We find that in this model, which is closely related to the BFSS model at finite temperature, thermal fluctuations can also source a scale-invariant spectrum of scalar and tensor fluctuations.
\end{abstract}

\tableofcontents

\section{Introduction}

Superstring theory is a promising candidate for a self-consistent unified theory of quantum gravity and particle physics.  From the perspective of a cosmologist, a promising aspect of string theory is that it can potentially describe the physics of the early universe where the Standard Model of Particles Physics and Einstein gravity are known to break down.  Specifically, in \cite{Brandenberger:1988aj}, an emergent scenario based on string theory (String Gas Cosmology) was proposed in which the universe emerges from a quasi-static phase, the Hagedorn phase, as a gas of strings at a temperature close to its limiting temperature, also known as the Hagedorn temperature \cite{Hagedorn:1965st}.  In this emergent phase, thermal fluctuations of the gas of strings lead to an almost scale-invariant spectrum of cosmological perturbations with a slight red tilt \cite{Nayeri:2005ck}, and of gravitational waves with a slight blue tilt \cite{Brandenberger:2006xi}. These thermal fluctuations provide a source for structure formation as the emergent phase transitions to the radiation-dominated phase of Standard Big Bang Cosmology. This model provides an interesting alternative to inflation, which is known to be hard to realize in string theory \cite{Obied:2018sgi}. Moreover, the String Gas Scenario yields a non-singular cosmology.

One problem of String Gas Cosmology, however, is that the background evolution of the universe near the Hagedorn temperature, which leads to the resolution of the singularity, is not well understood.  This is the case because string theory is usually considered at the perturbative level of an an effective field theory on a classical background space-time. Such description is known to break down at very high densities and curvatures, and in particular in the very early universe.  Therefore, to study the early universe in the high curvature regime, one should consider a non-perturbative formulation of string theory.  There have been many proposals for non-perturbative formulations of string theory, most of which rely on matrix models.  The idea behind these models is that certain system of $N \times N$ matrices can provide non-pertubative description of superstring theory in the large N limit.  There are two main proposals for matrix theory, the BFSS model \cite{Banks:1996vh} and the IKKT model \cite{Ishibashi:1996xs}.  In the BFSS model, the eigenvalue distribution of the matrices describe space and depend on a continuous parameter $t$ which plays the role of time.  In the IKKT model, space-time is fully described by large N matrices: one holds information about time and the others hold information about space\footnote{A review of this model will be provided in section \ref{sec:rev}}.  These models provide a non-perturbative description of M-theory and the Type IIB string respectively.

Recently, a novel emergent scenario was suggested which makes use of the BFSS model \cite{Brahma:2021tkh}.  In this scenario, the universe emerges in a thermal state described by the BFSS model at finite temperature.  While doing so, thermal fluctuations of the BFSS model lead to an almost scale-invariant spectrum of scalar and tensor perturbations, just like in the case of String Gas Cosmology.  The universe then transitions to the radiation dominated phase of Standard Big Bang Cosmology, and thermal perturbations source structure formation. Steps have also been made to understand the time evolution of the metric at early times in the universe in matrix theory \cite{Brahma:2022dsd}, which is something that was as hitherto out of reach of most perturbative approaches (See \cite{Brahma:2022ikl} for a review on these two topics).

Given the recent success of the BFSS model in explaining structure formation, we now suggest an alternative scenario, in which the IKKT model at finite temperature sources structure formation.  In the new scenario that we are proposing in this paper, the universe emerges from a non-perturbative phase of the supersymmetric IKKT model\footnote{This model is described by bosonic matrices which describe space-time, and fermionic matrices which describe space-time fermions.} at finite temperature, and thermal fluctuations source the spectra of scalar and tensor perturbations which are both (approximately) scale-invariant.  When space-time emerges, the universe transitions to the radiation dominated phase of Standard Big Bang Cosmology, and thermal perturbations source structure formation as was the case in the BFSS scenario.  

Our new early universe scenario makes use of the IKKT model at finite temperature which, to the knowledge of the authors, is a system that has not yet been studied in the literature.  Consequently, the first portion of the paper will be dedicated to defining a thermal state for the IKKT model, and studying its properties.  Our approach to defining a thermal state for the IKKT model will build on known properties of string theory, where strings at finite temperature can be studied by compactifying the Euclidean time component of the string target space on a circle where space-time fermions acquire suitable anti-periodic boundary conditions (see \cite{Atick:1988si} for relevant papers on this topic).  Since the IKKT model provides a non-perturbative description of the Type IIB string target space, we postulate that the IKKT model at finite temperature can be studied by compactifying the Euclidean time component of the matrices on a circle where the fermionic matrices acquire anti-periodic boundary conditions.  Following this prescription, we will evaluate the free energy and energy of the system, and study the thermal fluctuations which lead to structure formation.

%If the universe emerges in such a thermal state, then the spectrum of scalar and tensor perturbations can be shown to be scale invariant by using the usual theory of linear cosmological perturbations. \SB{[SB: Are we repeating these statements?]}

The present paper is structured as follows.  In section \ref{sec:rev}, we review some aspects of the IKKT matrix model and how space-time emerges from it.  In section \ref{sec:thermo}, we obtain the IKKT action at finite temperature by compactifying the Euclidean time component of the model on a thermal circle where fermions acquire boundary conditions. In section \ref{sec:eff_act}, we evaluate the free energy of the system at finite temperature in the low temperature regime.  Finally, in section \ref{sec:cosmo}, we show how space-time can emerge with a scale-invariant spectrum of scalar and tensor perturbations.

\section{Review of the matrix models and emergent space}
\label{sec:rev}

The IKKT model is a proposed non-perturbative formulation of Type IIB string theory.  The idea behind this model is to describe the world sheet of the Type IIB string using large N matrices.  To see how this can be done, let us start with the Green-Schwarz action of the Type IIB string:
\be
S_{GS} = - T\int d^2\sigma \left( \sqrt{-h} + 2i \epsilon^{ab} \partial_a X^\mu \bar{\psi} \Gamma_\mu \partial_b \psi \right)\, , \quad \quad
h_{ab} = \partial_a X^\mu \partial_{b} X_\mu \, .
\ee
Equivalently, the action above can be rewritten in the following "Schild" form
\be
S_{Schild} = - T \int d\sigma^2 \sqrt{g} \left(- \frac{\alpha}{4} \{ X^\mu , X^\nu\}^2  + \frac{i}{2} \bar{\psi} \Gamma^\mu \{ X_\mu , \psi\} + \beta \right) \, ,
\ee
where the Poisson brackets are defined by:
\be
\{ X^{\mu} , X^\nu \} = \frac{1}{\sqrt{g}} \epsilon^{ab} \partial_a X^\mu \partial_b X^\nu \, .
\ee
This new action depends on an auxiliary field $\sqrt{g}$, which satisfies the equation of motion
\be
\sqrt{g} = \sqrt{\frac{\alpha}{2 \beta}} \sqrt{- h} \quad , \quad h = \frac{1}{2} \{ X^\mu , X^\nu\}^2 
\ee
Substituting the result above in the Schild action gives us
\be
S_{Schild} = - T \int d^2 \sigma \left( \sqrt{2\alpha \beta} \sqrt{-h} + \frac{i}{2} \bar{\psi} \Gamma^\mu \{ X_\mu , \psi\} \right) \, ,
\ee
which is equivalent to the Nambu-Goto action of the Type IIB string provided that $2\alpha \beta = 1$ and that we normalize the fermions.  In the Schild formalism, the string dynamics is determined by the Poisson bracket $\{X^\mu , X^\nu\}$.  By analogy with quantum mechanics, the Schild action can be discretised by replacing the Poisson brackets by a commutator and the integral by a trace.
\be
\{\, ,\, \} \implies - i [\, ,\, ] \,,  \quad  \quad \int d\sigma^2 \sqrt{g} \implies \Tr \, .
\ee
In this case, the target space coordinates $X^\mu$ and associated fermions $\psi$\footnote{We will suppress the spinor indices here and everywhere below to avoid clutter.} are now described by large $N \times N$ matrices, and we obtain the IKKT model action
\be
S_{IKKT} = - \frac{T}{4} \alpha\, \Tr \, [X^\mu , X^\nu]^2 - \frac{T}{2} \Tr \, \bar{\psi} \,\Gamma^\mu\, [X_\mu , \psi]  - \beta \, \Tr 1 \, .
\ee
The last term in the action above is non-dynamical and can be neglected.  Hence, by convention, we will use the following form for the IKKT model action
\be
S_{IKKT} = - \frac{1}{4g^2} \Tr\, [A^\mu , A^\nu]^2 - \frac{1}{2g^2} \Tr\,  \bar{\psi}\, \Gamma^\mu\, [A_\mu , \psi] \, ,
\label{eq:IKKT_action1}
\ee
where we have defined $X^\mu \equiv A^\mu$.  The action above can also be obtained by dimensionally reducing the action of a 10-dimensional super Yang-Mills theory to a point. In all cases, the resulting action is invariant under SU(N) transformation.  In the action above, the $\mu, \nu$ is contracted with respect to the flat metric $\eta_{\mu \nu} = \text{diag}(+,-,...,-)$.  Hence, this model is often referred to as the Lorentzian IKKT model.  For the action to be supersymmetric in 10 dimensions, the fermions must be described by Majorana-Weyl spinors, which satisfy $\bar{\psi} = \psi \,C_{10}$.  Here, $C_{10}$ is the charge conjugation operator in ten dimensions, which we will take to satisfy $C_{10} \Gamma^{\mu} C_{10}^{-1} = - \Gamma^T$ and $C^{T}_{10} = - C_{10}$.

%Although the IKKT model is primarily a theory of strings, higher dimential structures, such as D-branes for example, are known to exist as solutions to the equations of motion related to the IKKT action.  Although these structures are sometimes hard to picture given the abstract nature of the large N matrices, some observables can be defined to describe their shape in time in space.  One of these observables is the moment of inertia tensor
%\be
%T^{\mu\nu} = \frac{1}{N} \Tr (A^\mu A^\nu) \, , 
%\ee
%which describes the extent eigenvalue distribution in various temporal or spacial directions.  The spacial elements of the moment of inertial tensor can be combined into the extent of space parameter:
%\be
%R^2 = \frac{1}{N} \Tr (A^i)^2 \, \, ,
%\ee
%which describes the overall extent of eigenvalue distribution in space.

\section{Compactification, SUSY breaking and thermodynamics}
\label{sec:thermo}

%As reviewed in the previous section, the IKKT model provides a matrix description of the Type IIB string worldsheet.  

In superstring theory, one can obtain a thermal state of the string by compactifying the time direction of a Euclidean target space on a torus together with suitable anti-periodic boundary conditions for the target space fermions.\footnote{This choice of compactification is closely related to the Scherk-Schwartz orbifolding procedure \cite{Scherk:1978ta}.}  This antiperiodic boundary condition for the fermions is required to break supersymmetry, which is expected to occur in any thermal system.  Given that the IKKT model describes the target space of the Type IIB string, one should expect to be able to construct a thermal state of the IKKT model by compactifying the matrices on such a torus.  The precise sense in which this can be done has already been explored by Taylor \cite{Taylor:1996ik} (symmetric boundary conditions for the fermions), Banks \cite{Banks:1999tr} (anti-symmetric boundary conditions for the fermions) and many other authors (\cite{Connes:1997cr}, \cite{Schreivogl:2013qza} and more).  In the present section, we will build on the approach of these authors to compactify the IKKT model on a Eucledian time circle, and study the thermodynamic properties of this system.

%The idea is to define an operator that translate the matrices $A^\mu$ around the time circle, and use this operator to define constraints on $A^\mu$ that describe the boundary conditions.  Under these constraints, the matrices acquire winding modes as a consequence of the compactification.

\subsection{Compactification and thermodynamics}

\begin{figure}[h]
	\centering
        \vspace{0.5cm}
	\includegraphics[width=12.5cm]{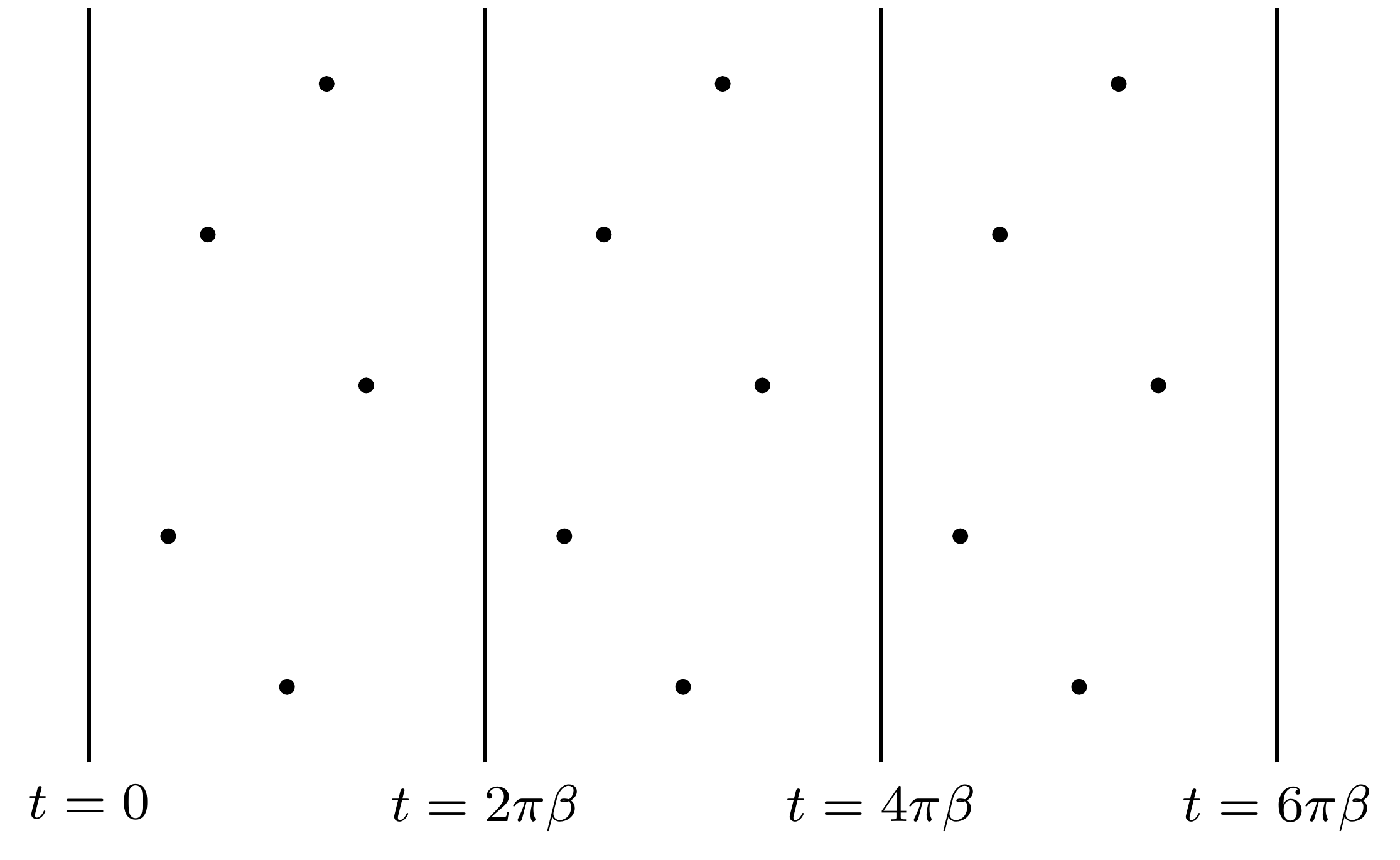}
	\caption{Sketch of mirror D-instantons in the duplicated 
        fundamental regions along the Euclidean time direction.  Each fundamental region, where the distribution of D-instantons is described by $A^\mu_0$, has an infinite number of mirror regions located at a distance $t = 2 \pi \beta n$ from each other along the $A^0$ direction, where $n$ is an integer.}
	\label{fig:mirror}
\end{figure}

Let us start by Euclideanising the IKKT action.  For a Lorentzian metric in the mostly minus signature ($+,-,...,-$), this can be done by the changes $A^0 \rightarrow iA^0$ and $\Gamma^i \rightarrow i \Gamma^i$ in the IKKT action (Equation \ref{eq:IKKT_action1}).  We obtain
\be
S_{IKKT} = - \frac{1}{4g^2} \Tr [A^\mu , A^\nu]^2 - \frac{i}{2g^2}\Tr \left( \psi C_{10} \Gamma^\mu [A_\mu , \psi] \right) \, ,
\label{eq:IKKT_action}
\ee
where the indices are now contracted with respect to the Euclidean metric $g_{\mu \nu} = \delta_{\mu \nu}$.  We will then compactify the time direction in the Euclidean IKKT model on a circle where the fermions acquire anti-periodic boundary conditions. To do this, let us presume the existence of an operator $U$ which translates the matrices by an amount $2 \pi \beta$ in the $A^0$ direction, where $\beta = 1/T$ is identified as the inverse temperature of the system.  To impose the desired boundary conditions, we want this operator and the matrices $A^\mu$ and $\psi$ to satisfy:
\begin{align}
U^{-1} A^0 U & = A^0 + 2\pi \beta \label{eq:cons_A^0} \\
U^{-1} A^i U & = A^i \label{eq:cons_A^i} \\
U^{-1} \psi U & = - \psi \, . \label{eq:cons_psi}
\end{align}
In other words, $A^i$ and $\psi$ must have periodic and anti-periodic boundary conditions respectively, and $A^0$ must be periodic up to the circumference $2\pi\beta$ of the Euclidean time circle.  These constraint equations can be solved by using operators $q$ and $p$ that satisfy the Heisenberg commutation relations $[q,p] = i$\footnote{Note that contrary to some papers in the literature, we will not assume that $q$ is an integer.  Here, $q$ can take any value on the real axis just like in quantum mechanics.  Let us remember that this is something we are allowed to do if $A^\mu$ and $\psi$ are sufficiently large matrices, which we will be assuming here.}.  Let us consider the unitary operator
\be
U = 1 \otimes e^{-i 2\pi q} e^{-ip} \, ,
\label{eq:anz1}
\ee
which translates a state a eigenvector $|q\rangle$ of the operator $q$ to $|q+1\rangle$ up to a phase $e^{-i 2\pi q}$, and assume $A^i$ and $\psi$ take the form
\be
A^i = \sum_{n} A_n^i \otimes e^{i n p} \quad ,  \quad \psi = \sum_r \psi_r \otimes e^{i r p} \, .
\label{eq:anz2}
\ee
Here, we will assume the $A_n^i$'s and $\psi_r$'s are large $M \times M$ matrices that live in a Hilbert space  different from the one of $q$ and $p$, where we will take the latter to be described by large $N \times N$ matrices.  Hence, $A^i$ and $\psi$ will be described by large $MN \times MN$ matrices.  Using the ansatz of equation \ref{eq:anz1} and \ref{eq:anz2}, the constraint equations \ref{eq:cons_A^i} and \ref{eq:cons_psi} can easily be solved by invoking that $n$ and $r$ must be integers and half integers respectively.  This result follows from the fact that the phase factor $e^{-i 2 \pi q}$ behaves like a fermion parity operator $(-1)^{F}$ in the sense that it commutes with the bosonic matrices $A^\mu$ and anti-commutes with the fermionic matrices $\psi$ given the equation \ref{eq:anz2}.\footnote{Notice that this is where our approach departs from Tom Banks approach in \cite{Banks:1999tr}.  The matrices $A^\mu$ and $\psi$ we obtain are slightly different in our case.  However, they satisfy the desired properties given by equation \ref{eq:cons_A^0}, \ref{eq:cons_A^i} and \ref{eq:cons_psi}.}  Hence, we obtain the desired boundary conditions for the bosonic and fermionic matrices.  $A^0$ can be found in a similar way by adding a term of the form $2 \pi \beta q$, which transforms like $U^{-1} ( 2 \pi \beta q ) U = 2 \pi \beta$ under the action of $U$.  We obtain 
\be
A^0 = \sum_{n \in \mathbb{Z} } A_n^0 \otimes e^{i n p} + 1 \otimes 2 \pi \beta q \, .
\ee
The resulting matrices $A^\mu$ and $\psi$, in the $q$-basis, can be written as
\begin{align}
A^0_{q' q} & = \langle q'| A^0 |q \rangle = \sum_{n \in \mathbb{Z} } A_n^0 \otimes \delta_{q' - n , q} + 2 \pi \beta q \delta_{q' q} \\
A^i_{q' q} & = \langle q'| A^i |q \rangle = \sum_{n \in \mathbb{Z} } A_n^i \otimes \delta_{q' - n, q} \\
\psi_{q' q} & = \langle q'| \psi |q \rangle = \sum_{r \in \mathbb{Z} + 1/2} \psi_r \otimes \delta_{q' - r, q} \, .
\end{align}
The matrices above have (or at least partly in the case of $A^0$) some sort of Toeplitz structure in the sense that they satisfy $A^i_{q'  q} = A^i_{q'-q}$ when $q$ is an integer and $\psi_{q'q} = \psi_{q'-q}$ when $q$ is a half-integer.  This structure describes a system of mirror D-objects that live on the diagonal of $A^\mu_{q'q}$ and $\psi_{q'q}$.  The system of mirror objects is composed of a fundamental region, described by $A^\mu_0$ and $\phi_0$, which is translated by a distance $2 \pi \beta$ for each adjacent diagonal block (See Figure \ref{fig:mirror}).
The off-diagonal elements, on the other hand, are related to interactions between each fundamental region.

These off-diagonal elements can also be associated to modes of a string winding around a circle.  Independently of the q-basis chosen above, this can be seen by substituting the constrained values of $A^\mu$ and $\psi$ in the IKKT action.  We obtain
\begin{align}
	S_{IKKT} & = \frac{N}{2g^2} \Tr \left( \sum_{n} (2\pi \beta n)^2 A^i_{-n}A^i_{n} + i \sum_r 2\pi \beta r \psi_{-r} C_{10} \Gamma^0 \psi_{r} \right. \\
	& \left. + \sum_{nm}  4\pi\beta n [A^0_{-n-m}, A^i_{m}]^2A^i_{n} - \sum_{nml} [A^0_{-n-m-l} , A^i_l][A^0_m , A^i_n] \right. \\
	& \left. - \frac{1}{2} \sum_{nml} [A^i_{-n-m-l} , A^j_l][A^i_m , A^j_n] - i \sum_{rn} \psi_{-r-n} C_{10}\Gamma^0 [A^0_n , \psi_r] - i \sum_{rn} \psi_{-r-n} C_{10} \Gamma^i [A^i_n , \psi_r] \right) \, . \label{eq:IKKT_modes}
\end{align}
Here, we made use the identities
\be
[q , e^{inp}] = - n e^{inp}\, , \quad  \quad \Tr e^{i (n \pm m) p} = N \delta(n \pm m) \, ,
\ee
and traced over the $q,p$ degrees of freedom to simplify the sums over $n$ and $r$.  The compactified IKKT model action has the structure of a mode expansion, where the first two terms describe the contribution of bosonic and fermionic winding modes with frequencies $\omega_{n} = 2\pi\beta n$ ($n \in \mathbb{Z}$ and $n \in \mathbb{Z} + 1/2$ for fermions), and the other term describe interactions between the different modes.  In the decompactification limit $\beta \rightarrow \infty$, the non-zero modes become heavy and decouple from the system.  The system is then well approximated by the bosonic IKKT model action
\be
S_{IKKT} = - \frac{N}{4g^2} \Tr [A_0^\mu , A_0^\nu]^2 \, .
\label{eq:bosonic_IKKT}
\ee
In the mirror objects picture, this means the mirror region are far from each other that the interactions between each region can be neglected.  We recover N copies of the fundamental region, as reflected by the extra factor of N in the action.  It's also interesting to note that fermions are projected away in the decompactification limit, which is a consequence of supersymmetry breaking.

\subsection{Relation to the BFSS model at finite temperature}

Given that the IKKT model is a description of the Type IIB string, it should be possible to link Equation \ref{eq:IKKT_modes} to the mode expansion of the Euclidean BFSS model, which is related to the Type IIA string, using T-duality.  Let us start with the Euclidean BFSS action
\be
S_{BFSS} = \frac{1}{2g^2} \int d\tau \Tr \left( (D_\tau X^i)^2 - \frac{1}{2}[X^i , X^j]^2 +  \bar{\psi} \Gamma^{0} D_\tau \psi - i \bar{\psi} \Gamma^i [X^i , \psi]\right) \, ,
\ee
where the covariant derivative is defined by
\be
D_\tau X = \partial_t X - i[ X^0 , X] \, ,
\ee
and the fermions are Majorana-Weyl spinors that satisfy $\bar{\psi} = C \psi$.  Substituting the mode expansion 
\be
X^0 = \sum_{n} X^0_n e^{i \omega n t} \quad , \quad X^i = \sum_{n} X^i_n e^{i \omega n t} \quad , \quad \psi = \sum_{r} \psi_r e^{i \omega r t} 
\ee
in the BFSS action, we obtain
\begin{align}
	S_{BFSS} & = \frac{\beta}{2g^2} \Tr \left( \sum_{n} (2\pi T n)^2 X^i_{-n}X^i_{n} + i \sum_r 2\pi T r \psi_{-r} C_{10} \Gamma^0 \psi_{r} \right. \\ 
	& \left. + \sum_{nm} 4 \pi T n [X^0_{-n-m}, X^i_{m}]^2X^i_{n}
	 - \sum_{nml} [X^0_{-n-m-l} , X^i_l][X^0_m , X^i_n] \right. \\ 
	& \left. - \frac{1}{2} \sum_{nml} [X^i_{-n-m-l} , X^j_l][X^i_m , X^j_n]
	 - i \sum_{rn} \psi_{-r-n} C_{10} \Gamma^0 [X^0_n , \psi_r] - i \sum_{rn} \psi_{-r-n} C_{10} \Gamma^i [X^i_n , \psi_r] \right) \, ,
\end{align}
where we have used $\omega = 2\pi/T$. Letting $T \rightarrow 1/T$, we recover the mode expansion of the IKKT model (equation \ref{eq:IKKT_modes}) up to a normalization of the gauge coupling $g^2$.  This implies that the thermodynamics of the IKKT and BFSS are related by T-duality.  For example, the high-temperature limit ($T \rightarrow \infty$) will be related to the low-temperature limit ($T \rightarrow 0$) of the IKKT model.  For the BFSS model, the high-temperature limit is a perturbative limit.  Similarly, the low-temperature limit of the IKKT model will also be a perturbative limit, which we will explore in the following section.

Notice also that the diagonal elements of the compact IKKT action become the Matsubara frequencies of the BFSS model under T-duality. 
 This property is a consequence of the Toeplitz structure of the compactified $A^\mu$ and $\psi$ matrices, which was recently utilized to rewrite quantum field theories compactified on a Toroidal space-time in terms of Toeplitz matrices \cite{Yargic:2022ycw}.

\section{Free energy of the IKKT model at finite temperature}
\label{sec:eff_act}

In the IKKT model, space-time time emerges from the bosonic $A^\mu$ matrices.  As we saw in the previous section, this picture becomes different once the IKKT model is compactified.  Instead of describing a single system, the IKKT describes an infinite number of copies of the same system that interact more and more with each other as we increase the temperature.  In the zero temperature limit, one recovers N copies of the fundamental region which are far away from each other, and hence do not interact with each other.  To understand the thermodynamics of this system, we will treat the low temperature limit of the IKKT model in the same way as was done in \cite{Kawahara:2007ib} for the high temperature limit of the BFSS model.  We will integrate out the interactions (non-zero winding modes) in order to obtain the effective free energy felt in the fundamental regions (zero modes) as a function of temperature.  This will later allow us to study thermodynamic properties of the fundamental regions (\textit{e.g.,} thermal fluctuations), which are relevant for cosmology.

\subsection{Gauge fixing and other considerations}

Before computing the free energy, there are some considerations we have to make.  First, let us choose an appropriate gauge fixing to evaluate the path integral.  To do this, note that the compactification procedure is equivalent to studying fluctuations of the matrices $A^\mu$ and $\psi$ around a background where $A^0 = 2\pi \beta q$, $A^i = 0$ and $\psi = 0$.  In other words, we are imposing
\begin{align}
	A^\mu & = X^\mu + \tilde{A}^\mu \\
	\psi & = \xi + \tilde{\psi} \, ,
\end{align}
where we choose
\be
X^0 = 2 \pi \beta q \quad , \quad X^i = 0 \quad , \quad \tilde{A}^\mu = \sum_n A^\mu_n e^{i n p}
\ee
\be
\xi  = 0 \quad , \quad \tilde{\psi} = \sum_r \psi_r e^{i r p} \, .
\ee
Here, the only difference is that we are also imposing that the fluctuation matrices $\tilde{A}^\mu$ and $\tilde{\psi}$ are symmetric and anti-symmetric under the action of the unitary operator in equation \ref{eq:anz1}.  Such expansions have been studied extensively in \cite{Ishibashi:1996xs} and \cite{Fayyazuddin:1997yf}.  In these cases, the appropriate gauge fixing condition is
\be
P_\mu A^\mu = 0 \, ,
\ee
where $P^\mu$ is the adjoint operator
\be
P_\mu Y = [X_\mu , Y]
\ee
associated to the background matrices $X^\mu$.  In the case at hand, imposing the gauge condition projects out the non-zero modes of $X^0$, giving us $X^0 = A^0_0$.  This gauge can be fixed by adding a ghost part
\be
S_{gh} = - \frac{1}{g^2} \Tr\left([X_\mu,\bar{c}][A^\mu,c] \right)
\ee
to the IKKT action.  Here, the ghost matrices $c$ will also be compact and hence take the form
\be
c = \sum_{n \in \mathbb{Z}} c_n e^{i n p}
\ee
Tracing out the $q$ and $p$ degrees of freedom in the ghost action, we obtain the following mode expansion
\be
S_{gh} = \frac{N}{g^2} \sum_{n} (2\pi \beta n)^2 \Tr (\bar{c}_n c_n) - \frac{N}{g^2} \sum_{n m} 2\pi \beta (n+m) \Tr ( \bar{c}_{n+m} [A^0_m , c_n] ) \, .
\ee
As a second consideration, we will assume the gamma matrices $\Gamma^\mu$ and the charge conjugation operator $C_{10}$ are in the following representation:
\begin{align}
\Gamma^0 & = 1_{16} \otimes \sigma_1 \\
\Gamma^i & = \gamma^i \otimes \sigma_2 \\
C_{10} & = C_9 \otimes i\sigma_2
\end{align}
Here, the $\gamma^i$'s are a set of nine-dimensional (Euclidean) gamma matrices which satisfy $\{ \gamma^i, \gamma^j \} = 2\delta_{ij}$, and $C_9$ is the associated charge conjugation matrix satisfying $C_9 \gamma^i C_9^{-1} = \gamma^{i T}$.  We will also choose $\gamma^i$ to be in the Majorana representation (where the nine $\gamma^i$ are taken to be real and symmetric), in which case the charge conjugation matrix takes the simple form $C_9 = 1_{16}$.  Finally, we will impose the following choice of Majorana-Weyl spinor:
\be
\psi = \phi \otimes
\begin{bmatrix}
	1 \\
	0 
\end{bmatrix} \, .
\ee
Here, $\phi$ is a sixteen-component Majorana fermion.

Given the considerations above, the compactified IKKT action, including the ghosts, can be rewritten as a sum of the zero mode action $S_0$, winding mode terms $S_{w}$ and interaction terms $S_{int}$ in the following way
\begin{align}
	S_{IKKT} & = S_0 + S_{w} + S_{int}\\
	S_0 & = - \frac{N}{4g^2} \Tr [A_0^\mu , A_0^\nu]^2 \\
	S_{w} & = \frac{N}{g^2} \Tr \left( \frac{1}{2} \sum_{n \not = 0} (2\pi n \beta)^2 A^0_{-n} A^0_{n} + \frac{1}{2} \sum_{n \not = 0} (2\pi n \beta)^2 A^i_{-n} A^i_{n} + \frac{i}{2} \sum_r 2 \pi r \beta \phi_{-r} \phi_r + \sum_{n} (2\pi n \beta)^2 \bar{c}_n c_n \right) \\
	S_{int} & = - \frac{N}{2g^2} \Tr \left( \sum_{n = m \not = 0} 4\pi n \beta [A^0_{-n-m} , A_{m}^i]^2 A^i_{n} + \sum_{n = m = l \not = 0} [A^0_{-n-m-l}, A^i_{l}][A^0_m , A^i_n] \right. \\
	& \left. \quad + \sum_{n=m=l \not= 0} \frac{1}{2} [A^i_{-n-m-l} , A^j_n][A^i_m , A^j_l] + i \sum_{rn}  \phi_{-r-n} [A^0_n , \phi_r] - \sum_{nr} \phi_{-r-n} \gamma^i [A^i_n , \phi_r] \right. \\
	& \left. \quad + \sum_{nm} 4\pi (n+m) \beta \bar{c}_{n+m} [A^0_m , c_n] \right)\, ,
\end{align}
where $\sum_{n = m \not= 0}$ and $\sum_{n=m=l \not= 0}$ respectively imply that the $m = n \not= 0$ and $m = n = l = 0$ terms are excluded from the sum.  Notice that we added a winding mode term for $A^0$ in the action.  We can do this because the gauge condition $P_\mu A^\mu = 0$ imposes that all $A^0_n$'s are zero when $n \not= 0$.  Hence, adding the winding mode term for $A^0$ is equivalent to adding zero in the action.  Since we recovered a winding mode term for $A^0$, it's useful to rewrite $S_{IKKT}$ in the more condensed form
\begin{align}
	S_{IKKT} & = S_0 + S_{w} + S_{int} \\
	S_0 & = - \frac{g^2}{4N} \Tr [A_0^\mu , A_0^\nu]^2 \\
	S_{w} & = \Tr \left( \frac{1}{2} \sum_{n \not = 0} (2\pi \beta n)^2 A^\mu_{-n} A^\mu_{n} + \frac{i}{2} \sum_r 2 \pi \beta r \phi_{-r} \phi_r + \sum_{n} (2\pi \beta n)^2 \bar{c}_n c_n \right) \\
	S_{int} & = - \frac{1}{2} \sqrt{\frac{g^2}{N}} \sum_{n = m \not = 0} 4\pi \beta n \Tr ( [A^0_{-n-m} , A_{m}^i]^2 A^i_{n} ) \\
	&  \quad - \frac{1}{4} \frac{g^2}{N} \sum_{n = m = l \not = 0} \Tr([A^\mu_{-n-m-l}, A^\nu_{l}][A^\mu_m , A^\nu_n]) - \frac{i}{2}  \sqrt{\frac{g^2}{N}} \sum_{nr} \Tr(\phi_{-r-n} [A^0_n , \phi_r]) \\
	& \quad + \frac{1}{2} \sqrt{\frac{g^2}{N}} \sum_{nr} \Tr(\phi_{-r-n} \gamma^i [A^i_n , \phi_r]) - \frac{1}{2} \sqrt{\frac{g^2}{N}} \sum_{nm} 4\pi (n+m) \Tr(\bar{c}_{n+m} [A^0_m , c_n]) \, ,
\end{align}
were we made the redefinitions $A^\mu_n \rightarrow \sqrt{g^2/N} A^\mu_n$, $\psi_r \rightarrow \sqrt{g^2/N} \psi_r$ and $\psi_r \rightarrow \sqrt{g^2/N} \psi_r$ to get rid of the $N$ and $g^2$ dependance in $S_w$.  The form above will be useful when evaluating the free energy.  As we can see from the action above, the SO(10) symmetry of the system is explicitly broken when the temperature of the system is non-zero.  The symmetry is restored when $T \rightarrow 0$ and the zero modes dominate the effective action.

\subsection{Free energy at leading order}

Let us now derive the free energy of the system at leading order.  We start with the partition function
\begin{align}
	Z & = \prod_{lrmn} \int D A_l^\mu D \phi_r D \bar{c}_m D c_n \, e^{-S_0-S_{w}-S_{int}} \\
	& = Z_0 Z_{w} \la e^{-S_{int}}  \ra  \, ,
\end{align}
of the compact IKKT action. As shown above, this partition function can be split into a contribution from the zero-mode partition function $Z_0$, the winding modes partition function $Z_{w}$ and the expectation value of the interaction terms $\la e^{-S_{int}} \ra$.  Here, $Z_0$, $Z_{w}$ and $\la . \ra$ are defined as. 
\be
Z_0 = \int D A^\mu_0 e^{-S_0} \quad , \quad Z_{w} = \prod_{l \not= 0}\prod_{rmn} \int D A_l^\mu D \phi_r D \bar{c}_m D c_n \, e^{-S_{w}}
\ee
\be
\la \, . \, \ra = \frac{1}{Z_{0} Z_{w}} \prod_{lrmn} \int D A_l^\mu D \phi_r D \bar{c}_m D c_n \, \, . \, \, e^{-S_0-S_{w}} 
\ee
Given the partition functions above, the free energy
\begin{align}
F & = - T\ln Z \\
& = -T \ln Z_0 - T \ln Z_{w} - T \ln \la e^{-S_{int}} \ra \, ,   
\end{align}
can be evaluated perturbatively.  At leading order in perturbation theory, only the first two terms contribute significantly to the free energy.  The first term, which depends on $\ln Z_0$, is not very interesting since $\ln Z_0$ does not depend on $\beta$.  Hence, we won't pay much attention to it.  For the second term, the contribution to the energy can be found by carrying out a series of Gaussian integrals.  We first split the winding modes partition function in a bosonic part $Z_b$, a fermionic part $Z_f$, and a ghost part $Z_{gh}$ in the following way:
\be
Z_{w} = Z_{b} Z_{f} Z_{gh} \, , 
\ee
\be
Z_{b} = \prod_{n \not= 0} \int D A_n^\mu \, e^{- \frac{1}{2} \Tr \left( \sum_{n \not = 0} (2\pi \beta n)^2 A^\mu_{-n}A^\mu_{n} \right)} \, ,
\ee
\be
Z_{f} = \prod_{r} \int D \phi_r \, e^{- i \frac{1}{2} \Tr \left( \sum_r 2 \pi \beta r \phi_{-r} \phi_r \right)} \, ,
\ee
\be
Z_{gh} = \prod_{nm} D \bar{c}_n D c_m \, e^{- \Tr \left( \sum_{n} (2\pi \beta n)^2 \bar{c}_n c_n \right)} \, .
\ee
The Gaussian integrals above can be carried out to obtain the expression below:
\be
Z_b = \left( \prod_{n \not = 0} (2\pi \beta n)^2  \right)^{-DM^2/2} \, ,
\ee
\be
Z_f = \left( \prod_{r} 2\pi \beta i r  \right)^{pM^2/2} \, ,
\ee
\be
Z_{gh} = \left( \prod_{n \not = 0} (2\pi \beta n)^2  \right)^{M^2} \, .
\ee
Here, $D$ is the number of space-time dimensions and $p$ is the dimension of the $\phi$ spinors, which are respectively $D=10$ and $p=16$ in the present case.  However, we will keep $D$ and $p$ arbitrary for the sake of generality.  The products above are manifestly divergent.  However, these divergences can be tamed using the identities 
\be
\prod_{n = 1}^{\infty} \left(\frac{2\pi n}{\alpha}\right)^{-2} = \frac{1}{\alpha} \quad , \quad \prod_{n = 1}^{\infty} \left(\frac{2\pi (n-1/2)}{\alpha}\right)^{2} = 2
\ee
found by zeta function regularisation (See Appendix \ref{ap:zeta}).  Using the identities above, we obtain
\be
Z_b = \beta^{DM^2/2} \, ,
\ee
\be
Z_f = 2^{pM^2/2} \, ,
\ee
\be
Z_{gh} = \beta^{-M^2} \, .
\ee
Including all the terms above, we obtain the following contribution to the free energy at leading order
\begin{align}
F_{leading}
 & = - T \ln Z_0 - T \ln Z_w \\
& =  -T \ln Z_0 - T M^2 (D-2) \ln \left(\beta\right) - \frac{1}{2} T pM^2 \ln 2 \, ,
\end{align}
and the following expression for the energy:
\be
E_{leading} = - \partial_\beta \ln Z_0 - \partial_\beta \ln Z_w = - M^2(D - 2) T \, .
\label{eq:E_w}
\ee
Here, $\ln Z_0$ is a constant that does not depend on the temperature of the system, and hence does not contribute to the energy.  Note that unlike for the BFSS model, the leading order contribution to the energy is negative.  This is a consequence of the fact that the Matsubara frequencies of the system are winding modes, and not Kaluza-Klein modes of a field compactified on a thermal circle like in thermal field theory.  The positive sign found for the BFSS model can be recovered by letting $T \rightarrow 1/T$ in the partition function, and computing the energy again using equation \ref{eq:E_w}.  Also note that, similar to the BFSS model (or all supersymmetric theories for that matter), the breaking of supersymmetry plays an important role in obtaining an non-vanishing contribution to the energy at leading order.  If supersymmetry was restored by giving the fermions periodic boundary conditions, the contribution from the bosonic sectior, the fermionic sectior and the ghosts would cancel giving $\ln Z_w = 0$ and $E_{leading} = 0$, and leaving only a contribution from $\ln Z_0$ to the free energy.

This is to be expected since, as we mentioned before, our chosen compactification is equivalent to perturbing the action around a background $X^\mu$ where $X^0 = 2 \pi \beta q$, $X^i = 0$ and $\xi = 0$, and imposing that the bosonic and fermionic fluctuations have periodic and anti-periodic boundary conditions respectively.  Such backgrounds describe a distribution of D-instantons that satisfy the BPS condition $F_{\mu\nu} = i [P_\mu , P_\nu] = 0$.  In this case, the one-loop partition function is known to vanish as a consequence of supersymmetry.  However, supersymmetry is broken here because of our choice of boundary conditions, so we obtain a non-vanishing contribution to the energy.

\subsection{Free energy at next to leading higher order}

The next to leading order terms in perturbation theory can be evaluated by expanding $\langle e^{-S_{int}} \rangle$ and evaluating the expectation values.  To obtain an effective description of the zero modes of the theory, we will evaluate the expectation values $\langle \langle \, . \, \rangle \rangle$ associated to the non-zero modes of theory in order to obtain an expression that depends on the expectation values $\la . \ra_0$ related to the zero modes in the theory.  The resulting expression will give us corrections to the zero-mode effective action.  Here, the expectation value $\langle \langle \, . \, \rangle \rangle$ with respect to the non-zero modes is defined by
\be
\la \la \, . \, \ra \ra = \frac{1}{Z_{w}} \prod_{l \not= 0}\prod_{rmn} \int D A_l^\mu D \phi_r D \bar{c}_m D c_n \, \, . \, \, e^{-S_{w}} \, .
\ee
and the expectation value $\la . \ra_0$ with respect to the zero modes $A^\mu_0$ is defined by
\be
\la \,. \, \ra_0 = \frac{1}{Z_0}  \int D A^\mu_0 e^{-S_0} \, .
\ee 
To evaluate the correction terms, it's useful to write down the two-point functions associated to the bosonic matrices, the fermionic matrices and the ghosts:
\begin{align}
\la \la (A^\mu_m)_{ab} (A^\nu_n)_{cd} \ra \ra & = \frac{\delta_{\mu \nu} \delta_{n+m,0} \delta_{ad} \delta_{bc}}{(2\pi \beta n)^2} \\
\la \la (\phi^\alpha_r)_{ab} (\phi^\beta_s)_{cd} \ra \ra & = \frac{ \delta_{\beta \alpha} \delta_{r+s,0} \delta_{ad} \delta_{bc}}{ 2 \pi \beta i r} \\
\la \la (\bar{c}_m)_{ab} (c_n)_{cd} \ra \ra & = \frac{\delta_{nm} \delta_{ad} \delta_{bc}}{(2\pi \beta n)^2} \, .
\end{align}
Here, $\alpha , \beta$ and $a, b, c, d$ are respectively spinor and matrix indices.  We will also separate the interaction part of the action into five different interaction terms in the following way:
\be
S_{int} = \sum_{p = 1}^5 V_p \, .
\ee
Here, each interaction term can be written as:
\begin{align}
V_1 & = - \frac{g^2}{4N} \sum_{n = m = l \not = 0} \Tr \left( [A^\mu_{-n-m-l}, A^\nu_{l}][A^\mu_m , A^\nu_n] \right) \, , \\
V_2 & = - \frac{1}{2} \sqrt{\frac{g^2}{N}} \sum_{n = m \not = 0} 4\pi \beta n \Tr \left( [A^0_{-n-m} , A_{m}^i] A^i_{n} \right) \, , \\
V_3 & = - \frac{i}{2} \sqrt{\frac{g^2}{N}} \sum_{nr} \Tr \left( \phi_{-r-n} [A^0_n , \phi_r] \right) \\
V_4 & =  \frac{1}{2} \sqrt{\frac{g^2}{N}} \sum_{nr} \Tr \left( \phi_{-r-n} \gamma^i [A^i_n , \phi_r] \right) \\
V_5 & = - \sqrt{\frac{g^2}{N}} \sum_{nm} 2\pi (n+m) \beta \Tr(\bar{c}_{n+m} [A_m^0 , c_n]) \, .
\end{align}
Using the interaction terms above, the corrections to the effective action at two-loop order can be written as
\be
\ln \la e^{- S_{int}} \ra = - \la V_1 \ra + \frac{1}{2} \la V_2^2 \ra + \frac{1}{2} \la V_3^2 \ra + \frac{1}{2} \la V_4^2 \ra + \frac{1}{2} \la V_5^2 \ra + ... \, ,
\ee
where each expectation value in the expansion above is given by
\begin{align}
\la V_1 \ra & = \frac{(D-1)}{12} \frac{M T^2 g^2}{N} \Tr \la A^\mu_0 A^\mu_0 \ra_0 + \mathcal{O}(T^4)  \\
\la V_2^2 \ra & = \frac{(D-1)}{3} \frac{M T^2 g^2}{N} \Tr \la A^0_0 A^0_0 \ra_0 + \frac{1}{6} \frac{M T^2 g^2}{N} \Tr \la A^i_0 A^i_0 \ra_0 + \mathcal{O}(T^4) \\
\la V_3^2 \ra & = - \frac{p}{4} \frac{M T^2 g^2}{N} \Tr \la A^0_0 A^0_0 \ra_0 + \mathcal{O}(T^4) \\
\la V_4^2 \ra & = \frac{p}{4} \frac{M T^2 g^2}{N} \Tr \la A^i_0 A^i_0 \ra_0 + \mathcal{O}(T^4) \\
\la V_5^2 \ra & = - \frac{1}{6} \frac{M T^2 g^2}{N} \Tr \la A^0_0 A^0_0 \ra_0 + \mathcal{O}(T^4) \, .
\end{align}
Here, we made use of the sum identities 
\be
\sum_{n \not = 0} \frac{1}{(2\pi n)^2} = \frac{1}{12} \quad , \quad \sum_{r} \frac{1}{(2\pi r)^2} = \frac{1}{4} 
\ee
to evaluate the expectation values.  Moreover, we only kept the terms to quadradic order in temperature, which provides the dominant next to leading order contribution.  Adding all the terms together and restoring the initial dimensions of the zero modes by letting $A^\mu_0 \rightarrow \sqrt{N/g^2} A^\mu_0$, we obtain the following expression for the corrections to the effective action at two-loop order:
\be
\ln \la e^{- S_{int}} \ra =	\left(\frac{D-2}{12} - \frac{p}{8}\right) M T^2 \left( \Tr \la A^0_0 A^0_0 \ra_0 -  \Tr \la A^i_0 A^i_0 \ra_0 \right) + \mathcal{O}(T^4) \, .
\ee
Here again, the contribution above is a consequence of broken supersymmetry.  If we were to restore supersymmetry by giving the fermions periodic boundary conditions, the $p/8$ prefactor in the expression above would get replaced by $p/24$, leading to a cancelation of the term above.

Note that the expression above can be explicitly related to the expectation value of the extent of space $\la R^2 \ra_0 = \frac{1}{M} \Tr \la A^i_0 A^i_0 \ra_0$ of the system and what we could define as the expectation value of the extent of time $\la T^2 \ra_0 = \frac{1}{M} \Tr \la A^0_0 A^0_0 \ra_0$.  Here, $\la R^2 \ra_0$ and $\la T^2 \ra_0$ should be respectively viewed as the characteristic size of space and duration of time in the fundamental regions.  Since $SO(10)$ symmetry is preserved at tree level, we expect the distribution of eigenvalues of $A^0$ to be similar to the distribution of eigenvalues of $A^i$.  Consequently, we expect that $\la T^2 \ra_0$ can be approximated by
\be
\la T^2 \ra_0 \approx \frac{1}{D-1} \la R^2 \ra_0 \, .
\ee
Substituting the expression above in the effective action, we obtain the following correction to the free energy
\be
F_{next} = - T \ln \la e^{- S_{int}} \ra =  \left(\frac{D-2}{12} - \frac{p}{8}\right)\frac{D-2}{D-1} M^2 T^3 \la R^2 \ra_0
\ee
and the following correction to the energy of the system
\be
E_{next} = - \partial_\beta \ln \la e^{- S_{int}} \ra = - 2 \left(\frac{D-2}{12} - \frac{p}{8}\right) \frac{D-2}{D-1} M^2 T^3 \la R^2 \ra_0 \, .
\ee

\section{Application to early universe cosmology}
\label{sec:cosmo}

In the last section, we studied the IKKT model at finite temperature and derived its free energy up to the next to leading order.  Let us now consider a scenario where the universe emerges in a thermal state described by the IKKT model at finite temperature.  Here, we will assume a 4 dimensional universe emerges and that thermal fluctuations are sourced by a four-dimensional version of the IKKT model ($D = p = 4$). 

If the theory of linear cosmological perturbations apply, we can show that the spectrum of scalar and tensor perturbations sourced by thermal fluctuations of the IKKT model is scale invariant following the prescription given in \cite{Nayeri:2005ck}.  To do this, let us assume space-time is described by the following longitudinal gauge metric
\be
ds^2 = (1+2\Phi) dt^2 - a(t)^2 [(1-2\Phi) \delta_{ij} + h_{ij}] dx^i dx^j 
\ee
where $\Phi$ is the relativistic generalisation of the Newtonian gravitational potential, $h_{ij}$ is a transverse traceless tensor which describes excitations of the metric due to gravitational waves, and $a(t)$ is the scale-factor of an arbitrary cosmological background.

According to linear cosmological perturbation theory, the amplitude of curvature fluctuations on a scale $k$ , where $k$ denotes the comoving wave number, is related to energy fluctuations on that scale via 
\be
\langle |\Phi(k)|^2 \rangle = 16 \pi^2 G^2 k^{-4} \langle \delta T^0_0(k) \delta T^0_0 (k) \rangle \, ,
\label{eq:scalar_amp}
\ee
where $T^\mu_\nu$ is the energy-momentum tensor of matter, and $G$ is Newton's gravitational constant.  Similarly, the amplitude $h(k)$ of tensor perturbations can be related to transverse pressure fluctuations via 
\be
\langle |h(k)|^2 \rangle = 16 \pi^2 G^2 k^{-4} \langle \delta T^i_j(k) \delta T^i_j (k) \rangle \quad , \quad i \not= j \, .
\label{eq:tensor_amp}
\ee
If matter is in a thermal state, then the amplitude of density and transverse pressure perturbations in a box of radius $R$ can be found from the finite temperature partition function of the system.  Specifically, since 
\be
\langle T^\mu_\nu \rangle = 2 \frac{g^{\mu \lambda}}{\sqrt{-g}} \frac{\partial \ln Z}{\partial g^{\nu \lambda}} \, ,
\ee
the fluctuations of the stress-energy tensor can be expressed as 
\begin{align}
\langle \delta T^\mu_\nu(k) \delta T^\sigma_\lambda (k) \rangle & \equiv \langle T^\mu_\nu T^\sigma_{\lambda} \rangle - \langle T^\mu_\nu \rangle \langle T^\sigma_{\lambda} \rangle \\
& = 2 \frac{g^{\mu \alpha}}{\sqrt{-g}} \frac{\partial}{\partial g^{\alpha \nu}} \left(\frac{g^{\sigma \delta}}{\sqrt{-g}} \frac{\partial \ln Z}{\partial g^{\delta \lambda }}\right) + 2 \frac{g^{\sigma \alpha}}{\sqrt{-g}} \frac{\partial}{\partial g^{\alpha \lambda}} \left(\frac{g^{\mu \delta}}{\sqrt{-g}} \frac{\partial \ln Z}{\partial g^{\delta \nu}}\right)
\end{align}
The expression above may seem complicated.  However, component-wise, it can be expressed in terms of rather simple thermodynamic observables. To obtain these expressions, we first move to the position space representation of the matter perturbation correlation functions using
\be
\langle \delta T^0_0(k) \delta T^0_0 (k) \rangle = k^{-3}\langle \delta T^0_0(k) \delta T^0_0 (k) \rangle_R \quad , \quad \langle \delta T^i_j(k) \delta T^i_j (k) \rangle = k^{-3} \langle \delta T^i_j(k) \delta T^i_j (k) \rangle_R \, ,
\label{eq:pos_space_corr}
\ee
where $\langle . \rangle_R$ denotes the expectation value in the position space representation.  Then, the energy density correlation function can be expressed in terms of the heat capacity 
\be
C_V = \left( \frac{\partial E}{\partial T}\right)_V
\ee
of the system in the following way
\be 
\langle \delta T^0_0(k) \delta T^0_0 (k) \rangle_R = \la \delta \rho^2 \ra_R = \la \rho^2 \ra_R - \la \rho \ra_R^2 = \frac{T^2}{R^6} C_V \, .
\label{eq:dens_pert}
\ee
Similarly, the correlation function of the off-diagonal spatial components of the stress tensor 
\be
\langle \delta T^i_j(k) \delta T^i_j (k) \rangle_R = \langle (T^i_j) ^2 \rangle_R - \langle T^i_{j} \rangle_R^2 \quad , \quad i \not= j
\ee
can be related to transverse pressure perturbations in the following way
\be
\langle \delta T^i_j(k) \delta T^i_j (k) \rangle_R = \alpha \frac{T}{R^2} \frac{\partial \tilde{p}}{\partial R} \, ,
\label{eq:trans_pert}
\ee
where the pressure $\tilde{p}$ is related to the free energy $F$ of the system via
\be
\tilde{p} = - \frac{1}{V} \frac{\partial F}{\partial \ln R} \, .
\label{eq:press}
\ee
Here, $\alpha$ is a suppression factor of the transverse pressure perturbations compared to the diagonal pressure perturbations which satisfies $|\alpha| < 1$.

If we assume that matter behaves in the way described by the thermodynamics of the IKKT model, then it's possible to find the spectrum of cosmological perturbations sourced by the IKKT model from the effective action derived in Section \ref{sec:eff_act}.  Let us start with the power spectrum of scalar cosmological perturbations, which is defined by
\be
P_{\Phi}(k) = k^3 \langle |\Phi(k)|^2 \rangle \, .
\ee
Making use of equations \ref{eq:scalar_amp} and \ref{eq:pos_space_corr}, the power spectrum can then be related to the position space density fluctuations via
\be
P_{\Phi}(k) = 16 \pi^2 G^2 k^{-4} \langle \delta \rho^2(k) \rangle_R \, .
\ee
As we saw before, the density perturbations can be related to the heat capacity of the system via equation \ref{eq:dens_pert}.  This gives us
\be
P_{\Phi}(k) = 16 \pi^2 G^2 k^2 T^2 C_V (k R)^{-6} \, .
\ee
As we can see above, the spectrum of fluctuations can be scale invariant as long as $C_V \sim k^{-2}$.  This feature is known to arise in String Gas Cosmology, and in an emergent scenario involving the BFSS model as shown recently.  Indeed, we will now show that this feature also arises when considering the thermodynamics of the IKKT model.  Using the expression for the energy derived in section \ref{sec:eff_act}, find the heat capacity
\be
C_V = - N^2 (D-2) + 6 \left(\frac{p}{8} - \frac{D-2}{12} \right) \frac{D-2}{D-1} M^2 T^2 \la R^2 \ra_0 \, .
\ee
The first term gives a contribution to the power spectrum proportional to $k^2$, which is subdominant on large scales.  The second term, however, gives us a scale-invariant contribution since $\la R^2 \ra_0 \sim k^{-2}$.  Putting everything together, the scale-invariant contribution to the power spectrum gives us
\be
P_{\Phi}(k) = 96 \pi^2 G^2 (k R)^{-4} \left(\frac{p}{8} - \frac{D-2}{12} \right) \frac{D-2}{D-1} M^2 T^4
\label{eq:scalar_power}
\ee
Similarly, one can evaluate the power spectrum of tensor fluctuations from the position space representation of transverse matter perturbations using
\be
P_{h}(k) = k^3 \langle |h(k)|^2 \rangle \, ,
\ee
which, using equations \ref{eq:tensor_amp} and \ref{eq:pos_space_corr}, can be related to the transverse matter fluctuations in the following way:
\be
P_h(k) = 16 \pi^2 G^2 k^{-4} \langle \delta T^i_j(k) \delta T^i_j (k) \rangle_R \, .
\ee
Making use of equations \ref{eq:trans_pert} and \ref{eq:press} and the free energy derived in section \ref{sec:eff_act}, we obtain
\be
\langle \delta T^i_j(k) \delta T^i_j (k) \rangle_R = 2 \alpha \left(\frac{D-2}{12} - \frac{p}{8} \right) \frac{D-2}{D-1} M^2 T^4 R^{-4} \, .
\ee
This gives us a contribution to the spectrum of cosmological perturbations of the form
\be
P_h(k) = 32 \pi^2 G^2 (kR)^{-4} \alpha \left(\frac{D-2}{12} - \frac{p}{8} \right) \frac{D-2}{D-1} M^2 T^4 \, ,
\label{eq:tensor_power}
\ee
which is also scale-invariant.  Note that for equation \ref{eq:tensor_power} to be positive, $\alpha$ must be a negative quantity. Hence, we expect the diagonal and off-diagonal pressure perturbations to have opposite sign in this system.

Comparing the results (\ref{eq:scalar_power}) and (\ref{eq:tensor_power}), we find that the tensor-to-scalar ratio is given by
\be
r = \frac{|\alpha|}{3} \, .
\ee
In order to be consistent with the current observational bound on r, the value of $\alpha$ needs to be of the order $\mathcal{O}(10^{-1})$.  This is slightly better than for cosmological perturbations sourced from thermal fluctuations of the BFSS model, where we need $\alpha$ to be of the order $\mathcal{O}(10^{-2})$ to obtain a result consistent with observations.  Recall that although the transverse pressure perturbations are naturally smaller compared to the diagonal ones for thermal perturbations, they are not expected to get fine-tuned to be extremely small.  Hence, we expect $\alpha$ to be smaller than one, but not by many orders in magnitude.  In this sense, the IKKT model at finite temperature is more likely to source perturbations with the correct value of $r$ than the BFSS model at finite temperature.

\section{Conclusion}

In this paper, we began a study of matrix model thermodynamics and suggested an emergent scenario in which a non-singular cosmology emerges from a thermal system described by the IKKT model at finite temperature.  Inspired by string thermodynamics, we defined the IKKT model at finite temperature by compactifying its Euclidean time matrix on a circle where fermions acquire anti-periodic boundary conditions.  We found that if the early universe emerges in a thermal state of the IKKT model, then structure formation can be sourced by thermal fluctuations of the IKKT model at finite temperature, which yield scale invariant scalar and tensor perturbations.  

So far, we have assumed that the universe transitions to the radiation-dominated phase of Standard Big Bang cosmology after the emergent phase.  However, as discussed in section \ref{sec:thermo}, the low-temperature regime of the theory is dominated by the bosonic IKKT action.  Hence, it's possible that the late-time dynamics of the system can be described by known cosmological solutions of the bosonic IKKT model (e.g. \cite{Kim:2012mw}) with perhaps some thermal corrections.  This scenario could share interesting similarities with new numerical results \cite{Nishimura:2022alt} involving the IKKT model which suggests space-time can emerge accompanied by a transition from a Euclidean space-time metric to a Lorentzian space-time metric.  In our scenario, the Euclidean portion of our space-time could be described by a thermal state of the IKKT model which transitions into a Lorentzian space-time described by the bosonic IKKT model plus some thermal corrections.  The details of this late time transition and the subsequent time evolution of the universe needs to be worked out in future work.

Our early universe model shares many properties of String Gas Cosmology, where the evolution of the universe is driven by a gas of strings at finite temperature.  In comparison, our model relies on a matrix description of strings at finite temperature, which yields similar results.  Namely, we obtain an emergent 4-dimensional space-time from superstring theory, which lives in 10 dimensions, and thermal fluctuations in the emergent universe yield a scale-invariant spectrum of perturbations.  The details of the symmetry breaking have not been discussed here.  However, numerical evidence suggests that such a process is realizable in matrix theory \cite{Kim:2011cr}.  The details of this transition are currently under study (see \cite{Brahma:2022ifx} for progress on this topic, also see for an \cite{Steinacker:2021yxt} alternate approach to trying to solve this problem).  It could be interesting to figure out the connection between these results for the traditional world sheet description of the superstring, and its matrix description.  In doing so, perhaps matrix theory could give us a better understanding of String Gas Cosmology at early times.

In addition to it's resemblance with String Gas Cosmology, our new scenario  shares interesting similarities with another recent emergent scenario \cite{Agrawal:2020xek} where the early universe begins in a topological phase, which then transitions to Standard Big Bang Cosmology with a scale-invariant spectrum of cosmological perturbations.  Large N matrix models are known to have interesting topological properties \cite{tHooft:1973alw}. 
 Hence, it would be interesting to understand how (and, if) the two scenarios are related to one another.  This could be the another subject of further study.

\section*{Acknowledgements}

S.L. and S.B. wish to thank Simon Caron-Huot for useful discussions that inspired them to write this paper, and Robert Brandenberger and Keshav Dasgupta for useful comments on the draft.  S.L. is supported in part by the Fonds de recherche du Québec (FRQNT).  S.B. is supported in part by the Higgs Fellowship and by the STFC Consolidated Grant ``Particle Physics at the Higgs Centre''.  The research at McGill is supported in part by funds from NSERC and the Canada Research Chair program.  

For the purpose of open access, the authors have applied a Creative Commons Attribution (CC BY) license to any Author Accepted Manuscript version arising from this submission.

\appendix

\section*{Appendix}

\section{Zeta function regularisation}
\label{ap:zeta}

While evaluating the contribution of the winding modes to the path integral, we used the identities
\be
\prod_{n = 1}^{\infty} \left(\frac{2\pi n}{\alpha}\right)^{-2} = \frac{1}{\alpha} \quad , \quad \prod_{n = 1}^{\infty} \left(\frac{2\pi (n-1/2)}{\alpha}\right)^{2} = 2 \, 
\label{eq:id2}
\ee
to regulate various divergent products obtained by integrating over $A^\mu_n$, $\psi_r$, and $c_n$ at one loop order.  Such products show up frequently in quantum mechanics and quantum field theory, and can be tamed using Zeta function regularisation.  To obtain the first identity in equation \ref{eq:id2}, let us define the function
\be
\zeta_b(s) = \sum_{n=1}^{\infty} \left(\frac{2\pi n}{\alpha}\right)^{-2s} = \left(\frac{\alpha}{2\pi}\right)^{2s} \zeta(2s)
\ee
where
\be
\zeta(s) = \sum_{n=1}^{\infty} n^{-s}
\ee
is the Riemann Zeta function.  Taking the derivative of $\zeta_b(s)$ with respect to $s$, we obtain
\begin{align}
\zeta'_b(s) & = \sum_{n=1}^{\infty} \left(\frac{2\pi n}{\alpha}\right)^{-2s} \ln  \left(\frac{2\pi n}{\alpha}\right)^{-2} \\
& = 2 \left(\frac{\alpha}{2\pi}\right)^{2s} \left(\ln\left(\frac{\alpha}{2\pi}\right)\zeta(2s) + \zeta'(2s)\right) \, .
\end{align}
The expression above can be used to express the product $\prod_{n = 1}^{\infty} \left(\frac{2\pi n}{\alpha}\right)^{-2}$ as a function of the Zeta function and its derivative.  In the limit where $s = 0$, we obtain
\be
e^{\zeta'_b(0)} = \prod_{n=1}^{\infty} \left(\frac{2\pi n}{\alpha}\right)^{-2} = \left(\frac{\alpha}{2\pi}\right)^{2 \zeta(0)} e^{2 \zeta'(0)} \, .
\ee
Using the known values $\zeta(0) = - \frac{1}{2}$ and $\zeta'(0) = - \frac{1}{2} \ln (2\pi)$ of the Zeta function, we recover the desired identity:
\be
\prod_{n = 1}^{\infty} \left(\frac{2\pi n}{\alpha}\right)^{-2} = \frac{1}{\alpha} \, .
\ee
The section identity in equation \ref{eq:id2} can be obtained in a similar way.  We first define the function 
\be
\zeta_f(s) = \sum_{n=1}^{\infty} \left(\frac{2\pi(n-1/2)}{\alpha}\right)^{-s} = \left( \frac{\alpha}{2\pi}\right)^s \zeta(s,1/2) \, ,
\ee
where 
\be
\zeta(s,a) = \sum_{n = 0}^{\infty} \frac{1}{(n+a)^s}
\ee
is the Hurwitz Zeta function.  Then, we take the derivative of $\zeta_f(s)$ to obtain
\begin{align}
	\zeta'_f(s) & = \sum_{n=1}^{\infty} \left(\frac{2\pi (n-1/2)}{\alpha}\right)^{-s} \ln  \left(\frac{2\pi(n-1/2)}{\alpha}\right)^{-1} \\
	& = \left(\frac{\alpha}{2\pi}\right)^{s} \left(\ln\left(\frac{\alpha}{2\pi}\right)\zeta(s,1/2) + \zeta'(s,1/2)\right) \, .
\end{align}
Here again, the expression above can be related to the product $\prod_{n = 1}^{\infty} \left(\frac{2\pi (n-1/2)}{\alpha}\right)^{2}$ via the expression
\be
e^{-2\zeta'_f(0)} = \prod_{n=1}^{\infty} \left(\frac{2\pi (n-1/2)}{\alpha}\right)^2 = \left(\frac{\alpha}{2\pi}\right)^{-2 \zeta(0,1/2)} e^{-2 \zeta'(0,1/2)} \, .
\ee
Using the value $\zeta(0,1/2) = 0$ and $\zeta'(0,1/2) = - \frac{1}{2} \ln (2)$ of the Hurwitz Zeta function, we obtain  
\be
\prod_{n = 1}^{\infty} \left(\frac{2\pi (n-1/2)}{\alpha}\right)^{2} = 2 \, ,
\ee
as desired.

\end{document}